# Ferromagnetic Antenna and its Application to Generation and Detection of Gravitational Radiation

Fran De Aquino

Maranhao State University, Physics Department, S.Luis/MA, Brazil.
deaquino@uema.br

A new type of antenna, which we have called Ferromagnetic Antenna, has been considered for Generation and Detection of Gravitational Radiation. A simple experiment, in which gravitational radiation at 10 GHz can be emitted and received in laboratory, is presented.

## 1. INTRODUCTION

The *gravitational mass*, $m_g$, produces and responds to gravitational fields. It supplies the mass factors in Newton's famous inverse-square law of Gravitation $\left(F_{12} = Gm_{g1}m_{g2}/r_{12}^2\right)$. Inertial mass $m_i$ is the mass factor in Newton's 2nd Law of Motion $(F = m_i a)$.

Several experiments[1-6], have been carried out since Newton to try to establish a correlation between gravitational mass and inertial mass.

Some years ago J.F.Donoghue and B.R. Holstein[7] have shown that the renormalized mass for temperature $T = 0$ is expressed by $m_r = m + \delta m_0$ where $\delta m_0$ is the *temperature-independent mass shift*. In addition, for $T > 0$, mass renormalization leads to the following expressions for inertial and gravitational masses, respectively: $m_i = m + \delta m_0 + \delta m_\beta$ ; $m_g = m + \delta m_0 - \delta m_\beta$, where $\delta m_\beta$ is the *temperature-dependent mass shift* given by $\delta m_\beta = \pi \alpha T^2/3m_i$ .

This means that a particle's gravitational mass decreases with the increasing temperature and that only in absolute zero $(T = 0 K)$ are gravitational mass and inertial mass equivalent.

The expression of $\delta m_\beta$ obtained by Donoghue and Holstein refers solely to thermal radiation.

The general equation of correlation between $m_g$ and $m_i$ will be deduced here. Then we will show that the gravitational mass can be changed by means of Extreme-Low Frequency (ELF) electromagnetic radiation. Two experiments, using appropriated ELF radiation, has been carried out to test experimentally this equation. The experimental results are in agreement with the theoretical predictions. On the other hand, the detection of *negative* gravitational mass in both experiments suggest the possibility of dipole gravitational radiation. This fact is highly relevant because a gravitational wave transmitter can be built to generate detectable levels of gravitational radiation in the laboratory.

We have concluded from the experiments that detectable gravitational radiation fluxes can be emitted from *ferromagnetic materials* dipoles subjected to appropriated ELF electromagnetic radiation.

Here, we will present an experiment which involves the generation and detection of high-frequency gravitational waves based on this new technology.

# 2. THE CORRELATION $m_g/m_i$ .

In order to obtain the general expression of correlation between $m_g$ and $m_i$, we will start with the definition of *inertial* Hamiltonian, $H_i$, and *gravitational* Hamiltonian, $H_g$, i.e.,

$$H_i = c\sqrt{p_i^2 + m_i^2 c^2} + Q\varphi \qquad [1]$$

$$H_g = c\sqrt{p_g^2 + m_g^2 c^2} + Q\varphi \qquad [2]$$

where $m_i$ and $m_g$ are respectively, the inertial and gravitational *masses at rest* ; $p_i$ is the *inertial momentum* and $p_g$ the *gravitational momentum*; $Q$ is the electric charge and $\varphi$ is an electromagnetic potential.

A *momentum shift* , $\delta p$, on the particle, produces an *inertial Hamiltonian shift*, $\delta H$ ,given by

$$\delta H = \sqrt{(p_i + \delta p)^2 c^2 + m_i^2 c^4} - \sqrt{p_i^2 c^2 + m_i^2 c^4} \quad [3]$$

Fundamentally $\delta p$ is related to *absorption* or *emission* of energy.

In the general case of *absorption* and posterior *emission*, in which the particle acquires a $\delta p$ at the absorption and another $\delta p$ at the emission, the *total inertial Hamiltonian shift* is

$$\delta H = 2\left(\sqrt{(p_i + \delta p)^2 c^2 + m_i^2 c^4} - \sqrt{p_i^2 c^2 + m_i^2 c^4}\right) [4]$$

Note that $\delta H$ is always *positive*.

We now may define the correlation between $H_i$ and $H_g$ as follows

$$H_i = H_g + \delta H \qquad [5]$$

If $\delta H = 0$, $H_i = H_g$, i.e., $m_g = m_i$ .

In addition from the Eqs.[1] and [2], we can write:

$$H_i - H_g = \sqrt{p_i^2 c^2 + m_i^2 c^4} - \sqrt{p_g^2 c^2 + m_g^2 c^4} \quad [6]$$

For a particle at rest, $V = 0$; $p_i = p_g = 0$. Consequently, Eqs.[4] and [6] reduces to

$$\delta H = 2\left(\sqrt{\delta p^2 c^2 + m_i^2 c^4} - m_i c^2\right) \qquad [7]$$

and

$$H_i - H_g = (m_i - m_g)c^2 \qquad [8]$$

Substitution of Eqs[7] and [8] into Eq.[5] yields

$$(m_i - m_g)c^2 = 2\left(\sqrt{\delta p^2 c^2 + m_i^2 c^4} - m_i c^2\right)$$

From this equation we obtain

$$m_g = m_i - 2\left[\sqrt{1 + \left(\frac{\delta p}{m_i c}\right)^2} - 1\right]m_i \qquad [9]$$

This is the general expression of correlation between gravitational and inertial mass.

Note that the term inside the square bracket is always positive. Thus, except for *anti-matter* ($m_i < 0$), the second term on the right hand side of Eq.[9] *is always negative*.

In particular, we can look on the *momentum shift* $(\delta p)$ as due to absorption or emission of *electromagnetic energy* by the particle ( by means of *radiation* and/or by means of *Lorentz's force* upon the *charge* of the particle).

In the case of radiation ( photons with frequency $f = \omega/2\pi$ ), if $n$ is the number of absorbed (or radiated) photons by the particle of mass $m_i$, we can write

$$\delta p = n\hbar k_r = n\hbar\omega/(\omega/k_r) = U/(dz/dt) =$$
$$= U/v \qquad [10]$$

Where $k_r$ is the real part of the *propagation vector* $\vec{k}$ ; $k = |\vec{k}| = k_r + ik_i$ ; $U$ is the *electromagnetic energy absorbed or emitted by the particle* and $v$ is the *phase* velocity of the electromagnetic waves, given by:

$$v = \frac{dz}{dt} = \frac{\omega}{\kappa_r} = \frac{c}{\sqrt{\frac{\varepsilon_r\mu_r}{2}\left(\sqrt{1 + (\sigma/\omega\varepsilon)^2} + 1\right)}} \qquad [11]$$

$\varepsilon$ , $\mu$ and $\sigma$, are the electromagnetic characteristics of the medium in which the incident (or emitted) radiation is

propagating ( $\varepsilon = \varepsilon_r \varepsilon_0$ where $\varepsilon_r$ is the *relative electric permittivity* and $\varepsilon_0 = 8.854 \times 10^{-12} F/m$ ; $\mu = \mu_r \mu_0$ where $\mu_r$ is the *relative magnetic permeability* and $\mu_0 = 4\pi \times 10^{-7} H/m$ ). For an *atom* inside a body , the incident(or emitted) radiation on this atom will be propagating inside the body , and consequently , $\sigma = \sigma_{body}$ , $\varepsilon = \varepsilon_{body}$, $\mu = \mu_{body}$.

From the Eq.[10] follows that

$$\delta p = \frac{U}{v} = \frac{U}{c}\left(\frac{c}{v}\right) = \frac{U}{c}n_r \qquad [12]$$

where $n_r$ is the *index of refraction*, given by

$$n_r = \frac{c}{v} = \sqrt{\frac{\varepsilon_r \mu_r}{2}\left(\sqrt{1 + (\sigma/\omega\varepsilon)^2} + 1\right)} \qquad [13]$$

$c$ is the speed in a vacuum and $v$ is the speed in the medium.

By the substitution of Eq.[12] into Eq.[9], we obtain

$$m_g = \left\{1 - 2\left[\sqrt{1 + \left(\frac{U}{m_i c^2}n_r\right)^2} - 1\right]\right\}m_i \qquad [14]$$

Substitution of $U = n\hbar\omega = nhf$ into Eq.[14], gives

$$m_g = m_i - 2\left\{\sqrt{1 + \left\{\frac{nhf}{m_i c^2}n_r\right\}^2} - 1\right\}m \qquad [15]$$

Light can be substantially slowed down or frozen completely by optically inducing a quantum interference in a *Bose-Einstein condensate* [8] .This means an enormous index of refraction at ~$10^{14}$Hz. If t he speed of light is reduced to <0.1m/s, the Eq.[15] tell us that the gravitational masses of the *atoms* of the Bose-Einstein condensate become *negative*.

Let us now consider the particular macroscopic case in which all the particles inside a body have the same mass $m_i$. If $N/S$ is the average density of absorbed (or emitted) photons by the body (number of photons across the area unit), and $a$ is the area of the surface of each particle of mass $m_i$, then according to the *Statistical Mechanics*, the calculation of $n$ can be made based on the well-known method of *Probability of a Distribution* . The result is

$$n = \frac{N}{S}a \qquad [16]$$

Obviously the power $P$ of the incident radiation ( photons with frequency $f$ ), must be $P = Nhf/\Delta t = Nhf^2$, thus we can write $N = P/hf^2$ . Substitution of $N$ into Eq.[16] gives

$$n = \frac{a}{hf^2}\left(\frac{P}{S}\right) = \frac{a}{hf^2}D \qquad [17]$$

where $D$ is the *power density* of the absorbed ( or emitted) radiation. Thus Eq.[15] can be rewritten in the following form:

$$m_g = m_i - 2\left\{\sqrt{1 + \left\{\frac{aD}{m_i cvf}\right\}^2} - 1\right\}m \qquad [18]$$

For $\sigma \gg \omega\varepsilon$ Eq.[11] reduces to

$$v = \sqrt{\frac{4\pi f}{\mu\sigma}} \qquad [19]$$

By substitution of Eq.[19] into Eq.[18] we obtain

$$m_g = m_i - 2\left\{\sqrt{1 + \left\{\frac{aD}{m_i c}\sqrt{\frac{\mu\sigma}{4\pi f^3}}\right\}^2} - 1\right\}m_i \qquad [20]$$

This equation shows that, *elementary particles* (mainly electrons), *atoms* or *molecules* can have their *gravitational masses* strongly reduced by means of ELF radiation.

It is important to note that in the equation above, *ferromagnetic materials* with very-high $\mu$ request smaller value of $D$ .

One can easily show that the Eq.[18] is general for all types of wave. This means that *sound waves* can also change the gravitational mass of a particle. In this case, $v$, in the Eq.[18], is the speed of sound in the medium; $f$ is the frequency of the sound waves and $D$ the power density ( or intensity ) of the sound radiation. We then conclude from the Eq.[18] that elementary particles, atoms, etc., can have their gravitational masses strongly reduced by means of ELF *sound* radiation. For example, if $D \approx 1 \ W/m^2 \ \left(120\,dB\right)$ and $f = 1 \ \mu Hz$ .

## 3. EXPERIMENTAL TESTS

In order to check Eq.[20] experimentally, it was built an apparatus ( System H ) presented in Fig.1. Basically, a 9.9mHz Transmitter coupled to a special antenna.

The antenna in Fig.1 is a *half-wave dipole, encapsulated by a iron sphere* (purified iron, 99.95% Fe; $\mu_i = 5,000\mu_0$ ; $\sigma_i = 1.03 \times 10^7 \, S/m$ ). We will check the effects of the ELF radiation upon the gravitational mass of the ferromagnetic material (iron sphere) surrounding the antenna.

The *radiation resistance* of the antenna for a frequency $\omega = 2\pi f$ , can be written as follows [9]

$$R_r = \frac{\omega \mu_i \beta_i}{6\pi} \Delta z^2 \qquad [21]$$

where $\Delta z$ is the length of the dipole and

$$\beta_i = \omega \sqrt{\frac{\varepsilon_i \mu_i}{2}\left(\sqrt{1 + \left(\sigma_i / \omega \varepsilon_i\right)^2} + 1\right)} =$$

$$= \frac{\omega}{c}\sqrt{\frac{\varepsilon_{ri} \mu_{ri}}{2}\left(\sqrt{1 + \left(\sigma_i / \omega \varepsilon_i\right)^2} + 1\right)} =$$

$$= \frac{\omega}{c}(n_r) = \frac{\omega}{c}\left(\frac{c}{v_i}\right) = \frac{\omega}{v_i} \qquad [22]$$

where $v_i$ is the velocity of the radiation through the iron.

Substituting [22] into [21] gives

$$R_r = \frac{2\pi}{3}\left(\frac{\mu_i}{v_i}\right)\left(\Delta z f\right)^2 \qquad [23]$$

Note that when the medium surrounding the dipole is *air* and $\omega \gg \sigma/\varepsilon$ , $\beta \cong \omega\sqrt{\varepsilon_0 \mu_0}$ , $v \cong c$ and $R_r$ reduces to the well-know expression $R_r \cong \left(\Delta z \omega\right)^2 / 6\pi\varepsilon_0 c^3$ .

Here, due to $\sigma_i \gg \omega\varepsilon_i$ , $v_i$ is given by the Eq.[19]. Then Eq.[23] can be rewritten in the following form

$$R_r = \left(\Delta z\right)^2 \sqrt{\left(\frac{\pi}{9}\right)\sigma_i \mu_i^3 f^3} \qquad [24]$$

The *ohmic resistance* of the dipole is [10]

$$R_{ohmic} \cong \frac{\Delta z}{2\pi r_0} R_S \qquad [25]$$

where $r_0$ is the radius of the cross section of the dipole, and $R_S$ is the *surface resistance* ,

$$R_S = \sqrt{\frac{\omega\mu_{dipole}}{2\sigma_{dipole}}} \qquad [26]$$

Thus,

$$R_{ohmic} \cong \frac{\Delta z}{r_0}\sqrt{\frac{\mu_{dipole} f}{4\pi\sigma_{dipole}}} \qquad [27]$$

Where $\mu_{dipole} = \mu_{copper} \cong \mu_0$ and $\sigma_{dipole} = \sigma_{copper} = 5.8 \times 10^7 \, S/m$ .

The *radiated power* for an *effective* (*rms*) current $I$ is then $P = R_r I^2$ and consequently

$$D = \frac{P}{S} = \frac{\left(\Delta z I\right)^2}{S}\sqrt{\left(\frac{\pi}{9}\right)\sigma_i \mu_i^3 f^3} \qquad [28]$$

where $S$ is the *effective* area. It can be easily shown that $S$ is the outer area of the iron sphere ( Fig.1), i.e., $S = 4\pi r_{outer}^2 = 0.19 m^2$ .

The iron surrounding the dipole increases its inductance $L$. However, for series RLC circuit the *resonance frequency* is $f_r = 1/2\pi\sqrt{LC}$ , then when $f = f_r$ ,

$$X_L - X_C = 2\pi f_r L - \frac{1}{2\pi f_r C} = \sqrt{\frac{L}{C}} - \sqrt{\frac{L}{C}} = 0.$$

Consequently, the impedance of the antenna, $Z_{ant}$, becomes *purely resistive*, i.e.,

$$Z_{ant} = \sqrt{R_{ant}^2 + (X_L - X_C)^2} = R_{ant} = R_r + R_{ohmic}.$$

For $f = f_r = 9.9 mHz$ the length of the dipole is

$$\Delta z = \lambda/2 = v/2f = \sqrt{\pi/\mu_i \sigma_i f} = 0.070n = 70 mm.$$

Consequently, the *radiation resistance* $R_r$, according to Eq.[24], is $R_r = 4.56\mu\Omega$ and the *ohmic resistance*, for $r_0 = 13 mm$, according to Eq.[27], is $R_{ohmic} \cong 0.02\mu\Omega$. Thus, $Z_{ant} = R_r + R_{ohmic} = 4.58\mu\Omega$ and the *efficiency* of the antenna is $e = R_r/R_r + R_{ohmic} = 0.9956$ (99.56%).

The radiation of frequency $f = 9.9 mHz$ is totally absorbed by the iron along a critical thickness $\delta = 5z = 5/\sqrt{\pi f \mu_i \sigma_i} \cong 0.11 m = 110 mm$. Therefore, from the Fig.1 we conclude that the iron sphere will absorb practically all radiation emitted from the dipole. Indeed, the sphere has been designed with this purpose, and in such a manner that all their atoms should be reached by the radiation.

When the ELF radiation is absorbed by the iron atoms their gravitational masses, $m_{gi}$, are changed and, according to Eq.[20], become

$$m_{gi} = m_i - 2\left\{\sqrt{1 + \frac{\mu_i \sigma_i}{4\pi c^2 f^3}\left(\frac{a_i}{m_i}\right)^2 D^2} - 1\right\}m_i \quad [29]$$

Substitution of [28] into [29] yields

$$m_{gi} = m_i - 2\left\{\sqrt{1 + \left(\frac{\mu_i^2 \sigma_i}{6cS}\right)^2\left(\frac{a_i}{m_i}\right)^2 (\Delta z I)^4} - 1\right\}m_i \quad [30]$$

Note that the equation above doesn't depends on $f$.

Thus, assuming that the radius of the iron atom is $r_{iron} = 1.40\times10^{-10}m$; $a_i = 4\pi r_{iron}^2 = 2.46\times10^{-19}m^2$ and $m_i = 55.85(1.66\times10^{-27}kg) = 9.27\times10^{-26}kg$ then the Eq.[30] can be rewritten as follows

$$m_{gi} = m_i - 2\left(\sqrt{1 + 2.38\times10^{-4} I^4} - 1\right)m_i \quad [31]$$

The equation above shows that the gravitational masses of the iron atoms can be nullified for $I \cong 8.51 A$. Above this critical current, $m_{gi}$ becomes *negative*.

The Table 1 presents the experimental results obtained from the System H for the gravitational mass of the *iron sphere*, $m_{g(iron\ sphere)}$, as a function of the current $I$, for $m_{iron\ sphere} = 60.50 kg$ ( *inertial mass* of the iron sphere ). The values for $m_{g(iron\ sphere)}$, calculated by means of Eq.[31], are on that Table to be compared with those supplied by the experiment.

The experiment showed that the gravitational mass can be reduced, nullified or made *negative* by means of appropriated ELF electromagnetic radiation. The experimental results are in agreement with the theoretical predictions, calculated by means of Eq.[20].

In another previous experiment [11] we have built a system (called system-G) to test the Eq.[20]. In the system-G, a spiral antenna (*half-wave dipole*) was encapsulated by *iron powder* and the iron powder by a *annealed iron toroid*. We have checked the effects of the ELF electromagnetic radiation upon the gravitational mass of ferromagnetic toroid surrounding the ELF antenna, the results have been similar. However, the system-G works with very high electric currents(up to 300A) while the system-H, just up to 10A.

From the technical point of view, there are several applications to this discovery. Now we can build gravitational binaries, and to extract *energy* from any site of a gravitational field. The gravity control will be also very important to *Transportation* systems, and for *Telecommunication* too, as we will see soon after.

## 4.GRAVITATIONAL RADIATION

When the *gravitational field* of an object *changes*, the changes ripple outwards through space and take a finite time to reach other objects. These ripples are called *gravitational radiation* or *gravitational waves* .

The existence of gravitational waves follows from the General Theory of Relativity. In Einstein's theory of gravity the gravitational waves propagate at the speed of light.

Just as electromagnetic waves (EM), gravitational waves (GW) too carry energy and momentum from their sources. Unlike EM waves, however, there is no dipole radiation in Einstein's theory of gravity. The dominant channel of emission is quadrupolar. But the detection of *negative* gravitational mass suggest the possibility of dipole gravitational radiation.

This fact is highly relevant because now we can build a gravitational wave transmitter to generate detectable levels of gravitational radiation in the laboratory.

Let us consider an electric current $I_C$ through a conductor (*annealed iron wire* 99.98%Fe; $\mu = 350,000 \, \mu_0 ; \sigma = 1.03 \times 10^7 \, S/m$ ) submitted to ELF electromagnetic radiation with power density $D$ and frequency $f$ .

If the ELF electromagnetic radiation come from a half-wave *electric* dipole ( copper ) encapsulated by an *annealed iron* (purified iron, with the same characteristics of the *annealed iron wire*), the *radiation resistance* of the antenna for $\sigma >> \omega\varepsilon$ , according to Eq.[23], can be written as follows

$$R_r = (\Delta z)^2 \sqrt{\left(\frac{\pi}{9}\right)\sigma\mu^3 f^3} \qquad [32]$$

The *ohmic resistance* is

$$R_{ohmic} \cong \frac{\Delta z}{r_0} \sqrt{\frac{\mu_{dipole} f}{4\pi\sigma_{dipole}}} \qquad [33]$$

The *radiated power* for an *effective* (*rms*) current $I$ is then $P = R_r I^2$ and consequently, the *power density*, $D$ , of the emitted ELF radiation, is

$$D = \frac{P}{S} = \frac{(\Delta z I)^2}{S} \sqrt{\left(\frac{\pi}{9}\right)\sigma\mu^3 f^3} \qquad [34]$$

where $S$ is the area surround of the dipole.

For $f = 69.4 \mu Hz$ , the length of the dipole is

$$\Delta z = \lambda/2 = v/2f = \sqrt{\pi/\mu \, \sigma \, f} = 0.10m$$

The gravitational mass of the free-electrons into the wire ( electric current $I_C$ ) can be obtained by means of the substitution of [34] into [20],i.e.,

$$m_{ge} = m_e - 2\left\{\sqrt{1 + \left(\mu^4\sigma^2\left(\frac{a_e}{6m_e cS}\right)^2 (\Delta z I)^4\right)} - 1\right\}m_e \quad [35]$$

Substitution of numerical values into Eq.[35] leads to the following equation

$$m_{ge} = m_e - 2\left\{\sqrt{1 + \frac{\sim 10^{-6}}{S^2}I^4} - 1\right\}m_e \quad [36]$$

Thus, for $S \cong 0.1m^2$ and $I \cong 100A$ (current trough the ELF antenna) the gravitational mass of the free-electrons becomes

$$m_{ge} \cong -200 m_e$$

This means that they becomes "*heavy*" electrons.

Now consider a *half-wave* electric dipole whose elements are two cylinders of *annealed iron* (99.98%Fe; $\mu = 350,000\,\mu_0\,; \sigma = 1.03 \times 10^7\,S/m$) subjected to ELF radiation with frequency $f = 69.4\mu Hz$ (see Fig.2). The oscillation of the gravitational masses of the "*heavy*" electrons through this *ferromagnetic antenna* will generate a gravitational radiation flux very greater than the flux which is generated by the electrons at the "*normal*" state $(m_g \cong m_e)$. The *energy* flux carried by the emitted gravitational waves can be estimated by analogy to the oscillating *electric* dipole.

As we know, the *intensity* of the emitted *electromagnetic* radiation from an oscillating electric dipole ( i.e., the energy across the area unit by time unit in the direction of propagation) is given by [12]

$$F(\phi) = \frac{\pi^2 \Pi_0^2 f^4}{2c^3 \varepsilon r^2} sin^2 \phi \qquad [37]$$

The *electric dipole moment* , $\Pi = \Pi_0 sin\omega t$ , can be written as $qz$ , where $q$ is the oscillating electric charge, and $z = z_0 sin\omega t$ ; thus, one can substitute $\Pi_0$ by $qz_0$ , where $z_0$ is the amplitude of the oscillations of $z$ .

There are several ways to obtain the equivalent equation for the *intensity* of the emitted *gravitational* radiation from an *oscillating gravitational dipole*. The simplest way is merely the substitution of $\varepsilon$ (*electric permittivity*) by $\varepsilon_G = 1/16\pi G$ (*gravitoelectric permittivity* [13,14,15] ) and $q$ by $m_g$ (by analogy with electrodynamics, the *gravitoelectric dipole moment* can be written as $m_g z$ , where $m_g$ is the oscillating gravitational mass). Thus the *intensity* of the emitted *gravitational* radiation

from an oscillating *gravitational* dipole, $F_{gw}(\phi)$ , can be written as follows:

$$F_{gw}(\phi) = \frac{8\pi^3 G m_g^2 z_0^2 f_{gw}^4}{c^3 r^2} sin^2 \phi \qquad [38]$$

where $f_{gw}$ is the frequency of the gravitational radiation (equal to the frequency of the electric current through the ferromagnetic dipole).

Similarly to the electric dipole, the intensity of the emitted radiation from a gravitational dipole is maximum at the equatorial plane $(\phi = \pi/2)$ and zero at the oscillation direction $(\phi = 0)$ .

The gravitational mass $m_g$ in Eq.[38] refers to the *total* gravitational mass of the "*heavy*" electrons, given by

$$m_g = \left(10^{29}\,free-electrons/m^3\right) V_{ant} m_{ge}$$

where $V_{ant}$ is the volume of the antenna.

For the microwave *ferromagnetic antenna* in Fig.2, $V_{ant} = 7.4 \times 10^{-8}\,m^3$ and $m_{ge} \cong -200 m_e$ . This gives $m_g \cong 10^{-6}\,kg$ .

The amplitude of oscillations of the half-wave *gravitational* dipole is

$$z_0 = \lambda_{gw}/2 = \frac{c}{2 f_{gw}}$$

This means that to produce gravitational waves with frequency $f_{gw} = 10 GHz$ , the length of the ferromagnetic dipole, $z_0$ , must be equal to 1.5cm. By substitution of these values and $m_g \cong -10^{-6}\,kg$ into Eq.[38] we obtain

$$F_{gw}(\phi) \cong 10^{-9}\,\frac{sin^2 \phi}{r^2} \qquad [39]$$

At a distance $r = 1m$ from the dipole the maximum value of $F_{gw}(\phi)$ is

$$F_{gw}\left(\frac{\pi}{2}\right) \cong 10^{-9}\,W/m^2$$

For comparison, a gravitational radiation flux from astronomical source with frequency 1Hz and amplitude $h_{w0} \cong 10^{-22}$ ( the dimensionless amplitude of the gravitational waves of astronomical origin that could be detected on earth and with a frequency of about 1 kHz is between $10^{-17}$ and $10^{-22}$ ) has [16],

$$F_{gw} = \frac{1}{32\pi} \frac{c^3}{G} \left( \frac{dh_w}{dt} \right)^2 =$$

$$= 1.6 \times 10^{-5} \left( \frac{f}{100 Hz} \right)^2 \left( \frac{h_{w0}}{10^{-22}} \right)^2 \cong 10^{-9} W/m^2$$

As concerns detection of the gravitational radiation from dipole, there are many options. A similar gravitational dipole can also absorb energy from an incident gravitational wave. If a gravitational wave is incident on the gravitational dipole(receiver) in Fig.2(b) the masses of the "heavy" electrons will be driven into oscillation. The amplitude of the oscillations will be the same of the emitter, i.e., 1.5cm, and there will be an induced *electric* current $I^{'}$ through the ferromagnetic antenna of the receiver (see Fig.2(b)).

The weakness with which gravitational waves interact with matter is well-known. There is no significant scattering or absorption. It means that gravitational waves carry uncorrupted information even if they come from the most distant parts of the Universe. Consequently , a gravitational waves transceiver, based on the experiment presented in Fig.2, would allow us to communicate through the Earth, which, like all matter of the Universe, is transparent to gravitational waves. Furthermore, the receiver would allow us to directly observe for the first time the Cosmic Microwave Background in Gravitational Radiation, which would picture the Universe, just at the beginning of the Big Bang.

# REFERENCES

1. Eötvos, R. v. (1890), *Math. Natur. Ber. Ungarn*, **8**,65.

2. Zeeman, P. (1917), *Proc. Ned. Akad. Wet.*, **20**,542.

3. Eötvos, R. v., Pékar, D., Fekete, E. (1922) *Ann. Phys.*, **68**,11.

4. Dicke, R.H. (1963) *Experimental Relativity in "Relativity, Groups and Topology" (Les Houches Lectures)*, p. 185.

5. Roppl, P.G et. al. (1964) *Ann. Phys* (N.Y), **26**,442.

6. Braginskii, V.B, Panov, V.I (1971) *Zh. Eksp. Teor. Fiz*, **61**,873.

7. Donoghue, J.F, Holstein, B.R (1987) *European J. of Physics*, **8**,105.

8. M. M. Kash, V. A. Sautenkov, A. S. Zibrov, L. Hollberg, H. Welch, M. D. Lukin, Y. Rostovsev, E. S. Fry, and M. O. Scully, (1999) *Phys. Rev. Lett.* **82**, 5229 ; Z. Dutton, M. Budde, Ch. Slowe, and L. V. Hau, (2001) *Science* **293**, 663.

9. Stutzman, W.L, Thiele, G.A, *Antenna Theory and Design*. John Wiley & Sons, p.48.

10. Stutzman, W.L, Thiele, G.A, *Antenna Theory and Design*. John Wiley & Sons, p.49.

11. De Aquino, F. (2000) "*Possibility of Control of the Gravitational Mass by means of Extra-Low Frequencies Radiation*", Los Alamos National Laboratory, preprint gr-qc/0005107.

12. Alonso, M., Finn, E.J.(1972) *Física*, Ed. Edgard Blücher, p.297. Translation of the edition published by Addison-Wesley (1967).

13. Chiao, R. Y. (2002) "*Superconductors as quantum transducers and antennas for gravitational and electromagnetic radiation*", Los Alamos National Laboratory, preprint gr-qc/0204012 p.18.

14. Kraus,J.D. (1991) *IEEE Antennas and Propagation Magazine*, **33**, 21.

15. Landau, L. D. and Lifshitz, E.M. (1951) *The Classical Theory of Fields*, 1st edition, (Addison-Wesley) p..328 ; Forward, R. L. (1961) *Proc. IRE*, **49**, 892 ; Ciufolini, I. *et al.*, (1998) *Science*, 279, 2100.

16. Schultz, B.F. (2000) " *Gravitational Radiation*", Los Alamos National Laboratory, preprint gr-qc/0003069, p.5.

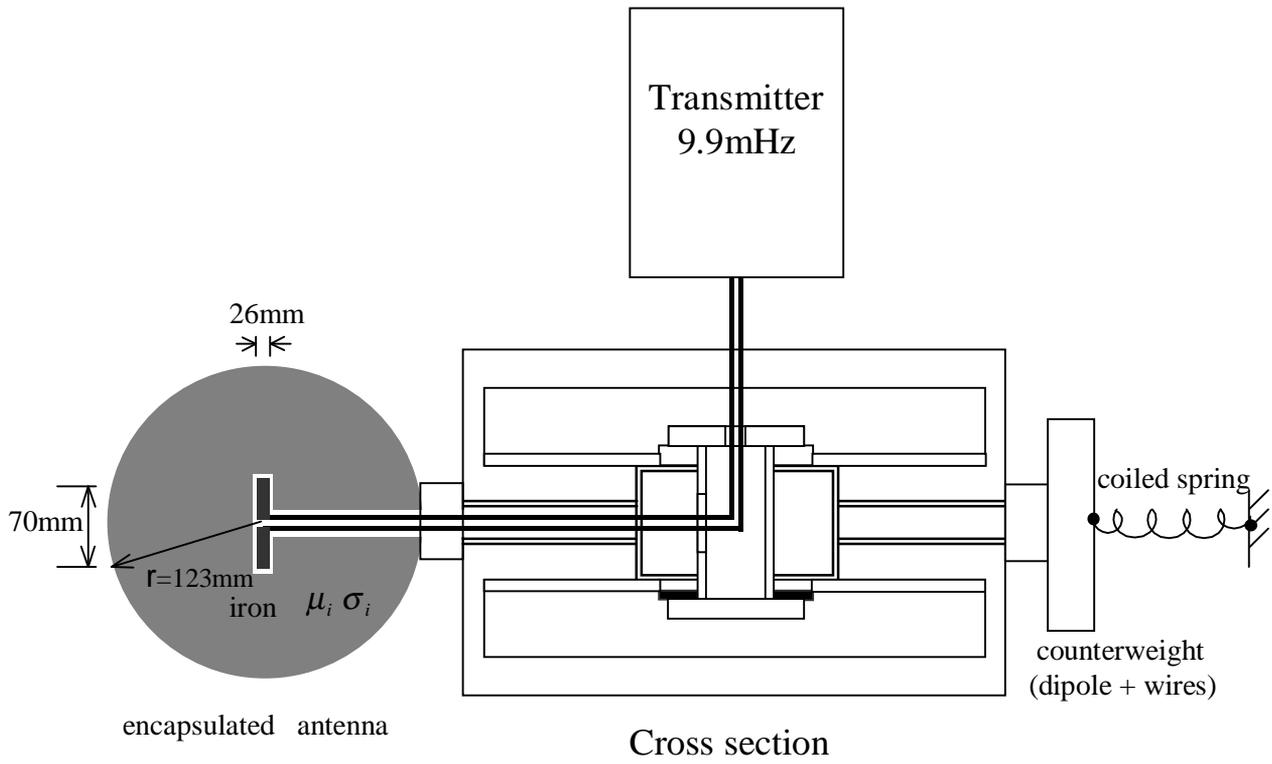

Transmitter
9.9mHz

26mm

70mm

r=123mm   iron   $\mu_i$  $\sigma_i$

encapsulated  antenna

coiled spring

counterweight
(dipole + wires)

Cross section

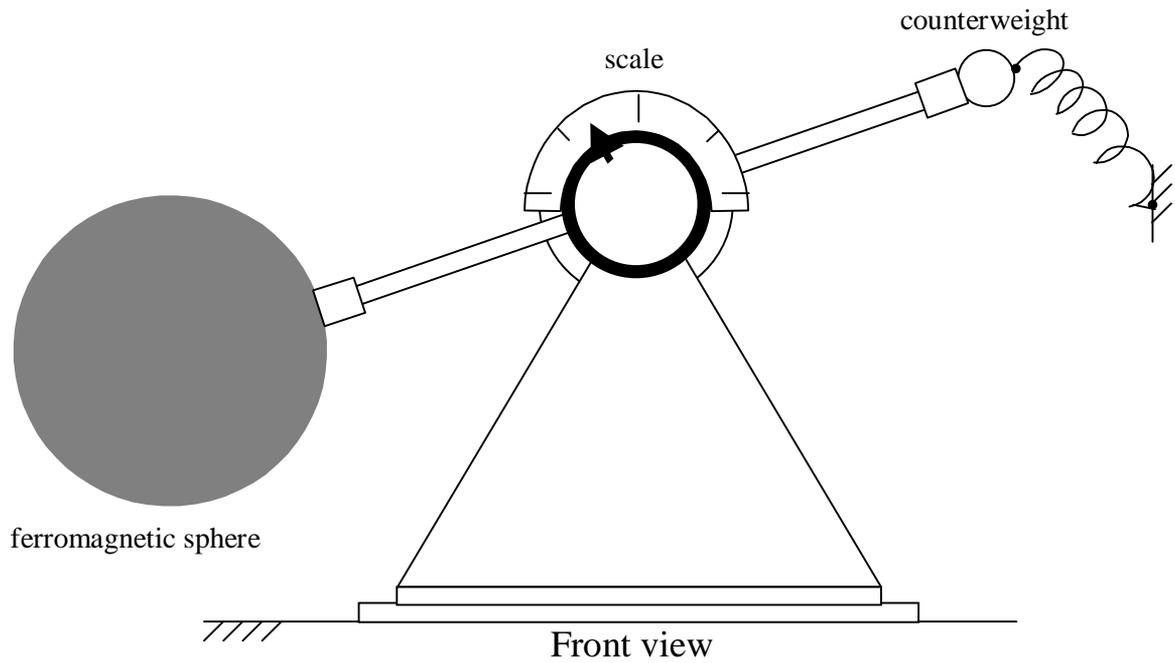

counterweight

scale

ferromagnetic sphere

Front view

Fig.1 - Schematic diagram of the System H

| $I$ (A) | $m_{g(iron\ sphere)}$ ( kg ) | |
|---|---|---|
| | theory | experimental |
| 0.00 | 60.50 | 60.5 |
| 1.00 | 60.48 | 60.(4) |
| 2.00 | 60.27 | 60.(3) |
| 3.00 | 59.34 | 59.(4) |
| 4.00 | 56.87 | 56.(9) |
| 5.00 | 51.81 | 51.(9) |
| 6.00 | 43.09 | 43.(1) |
| 7.00 | 29.82 | 29.(8) |
| 8.00 | 11.46 | 11.(5) |
| 8.51 | 0.0 | 0.(0) |
| 9.00 | -12.16 | -12.(1) |
| 10.00 | -40.95 | -40.(9) |

Table 1

Note: The *inertial mass* of the iron sphere is $m_{iron\ sphere} = 60.50 kg$

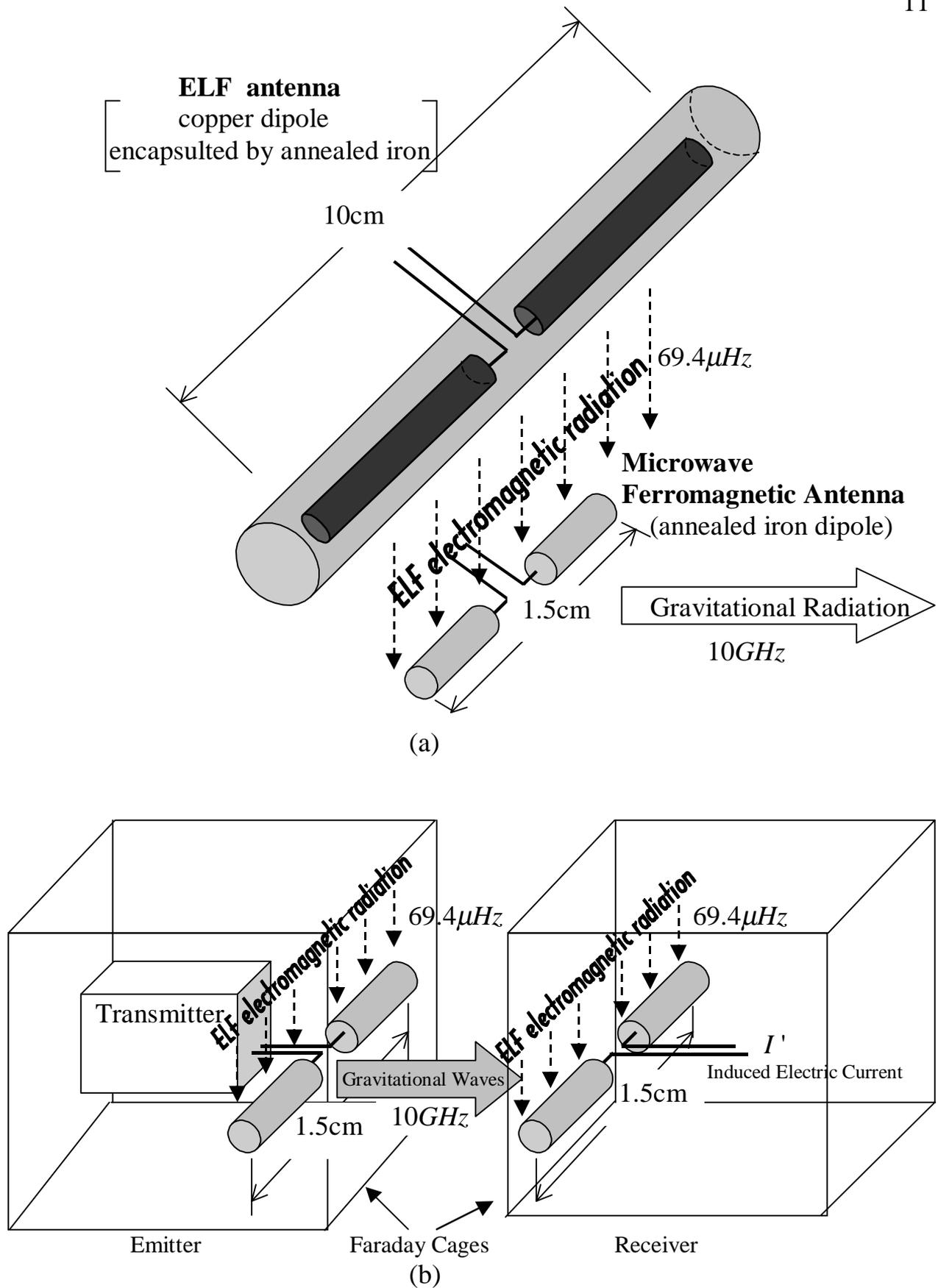

(a)

(b)

Fig.2 - Schematic diagram of the *ferromagnetic antennas* to produce and receive gravitational radiation.

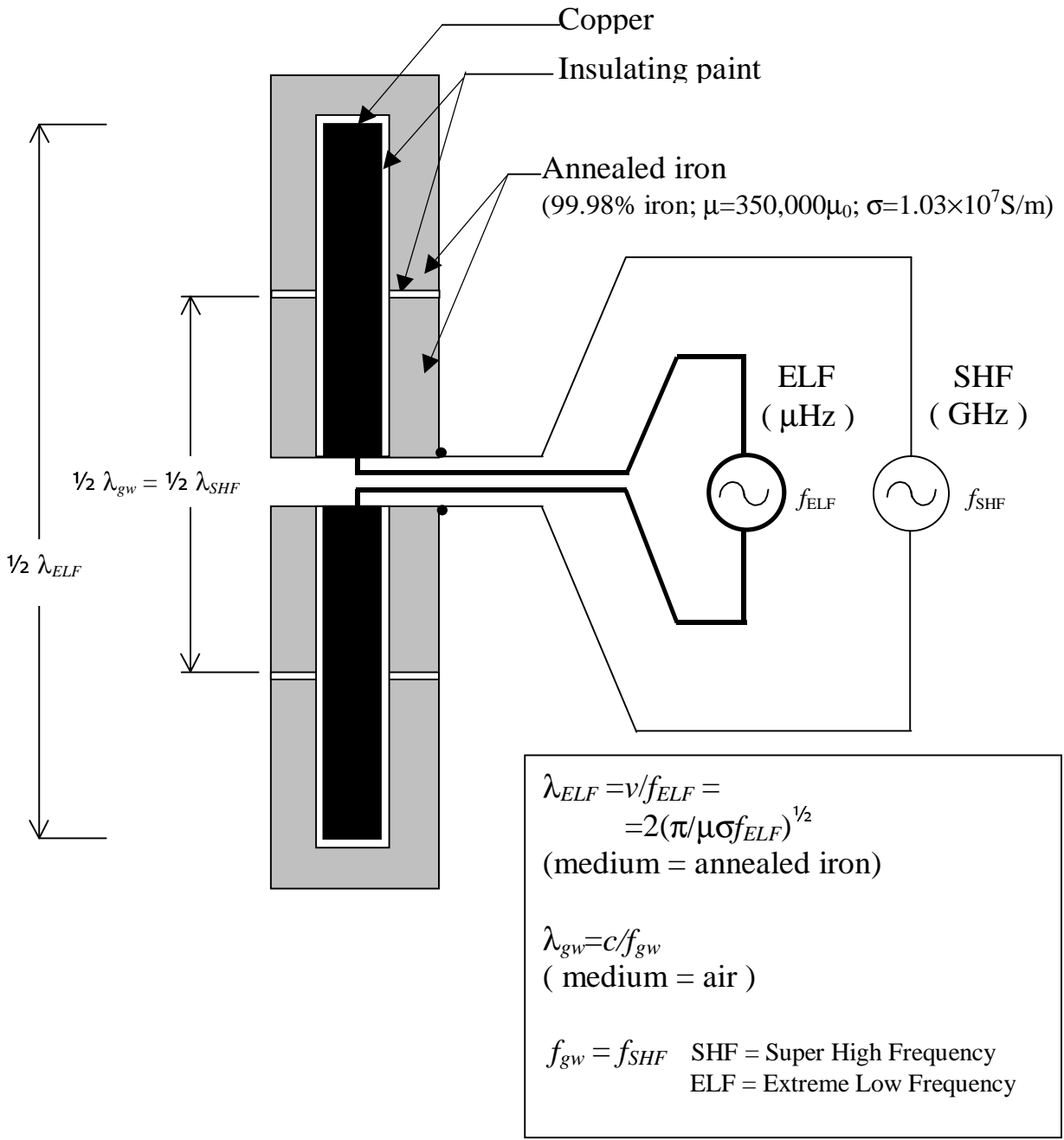

Fig.3 - An improved design for the GW antenna.